\documentclass[aps,prl,twocolumn,showpacs,floatfix]{revtex4}
\usepackage{amsmath,bm,epsfig}
\usepackage{latexsym}
\usepackage{amssymb}
\usepackage{color}
\usepackage{morefloats}
\usepackage{subfigure}
\usepackage{graphicx}

\begin{document}

\title{Reconstructing Links in Directed Networks from Noisy Dynamics}
\author{Emily S.C. Ching\footnote{ching@phy.cuhk.edu.hk}
 and H.C. Tam} \affiliation{Department of Physics, The Chinese
University of Hong Kong, Shatin, Hong Kong}

\date{\today}
\begin{abstract}
In this Letter, we address the longstanding challenge of how to
reconstruct links in directed networks from measurements, and
present a general method that makes use of a noise-induced
relation between network structure and both the time-lagged covariance of measurements taken
at two different times and the covariance of measurements
taken at the same time. For coupling functions that have
additional properties, we can further reconstruct the
weights of the links.

\end{abstract}

 \maketitle

The study of networks~\cite{Strogatz,AlbertBarabasi,Newman} has
emerged in many branches of science. Many systems of interest
consist of a large number of components that interact with each
other. These systems can be represented as networks with the
individual components being the nodes or vertices and the
interactions between two nodes being the links or edges that join
the nodes. The network structure depicting how the nodes are
linked is a crucial piece of information for us to understand the
behavior and function of the system that the network represents.
It is often difficult to directly measure the network structure
while the dynamics of individual nodes can be measured with
relative ease. This leads to the interesting question of how to
reconstruct the links of a network from the measurements of the
nodes. For most systems of interest, one node can affect the
dynamics of another node but its own dynamics is unaffected by the
latter. These systems are represented as directed networks with
directional links. In general, the strength of interaction can be
different so the links have different weights.

Most existing reconstruction methods apply only for undirected
networks in which interactions between two nodes are mutual.
 A commonly employed idea is
to infer links from correlation of measurements, with a higher
correlation interpreted as a higher probability of a
link~\cite{StuartScience2003,BrainPRL}. For undirected networks
with certain forms of diffusive coupling, it has been
shown~\cite{noise,PRE,PRErapid} that information of the network
structure is contained in the inverse of the covariance matrix and
not the covariance matrix itself. This explains why systems can be
strongly coupled but have weak pairwise
correlation~\cite{weak}. Other reconstruction methods either
assume the dynamics to be linear~\cite{YeungPNAS,SauerPRE} or
require additional information such as knowledge about nodal
dynamics~\cite{Yu,Shandilya,Levnajic,Consensus,SciRep2014} and
response dynamics to specific perturbations~\cite{Timme}. A number
of techniques have been proposed for the detection of directional
coupling from time series measurements but these techniques have limitations,
particularly when applied to real-world
problems~\cite{Smirnov2005,Smirnov2007}. Reconstructing links of
directed networks from measurements remains a big
challenge~\cite{Timme0,TimmeReview}.

In this Letter, we present a general method that reconstruct links
in a directed network subject to noise.  We show that information of the network
structure is contained in a noise-induced relation
between the time-lagged covariance of measurements taken at two
different times and the covariance of measurements taken at the
same time. For coupling functions that have additional properties,
we can further reconstruct the weights or relative coupling strength of the
links.

We study a network of $N$ nodes and each node is described by a state variable $x_i(t)$, $i=1, 2, \ldots, N$.
The dynamics of the nodes are governed by \begin{equation}
\frac{d{x}_i}{dt} = f_i(x_i) + \sum_{j\ne i} g_{ij} A_{ij}
h(x_i,x_j) + \eta_i  \ , \label{network} \ \
\end{equation}
where $f_i$ describes the intrinsic dynamics of node $i$ and
$A_{ij}$ are the elements of the adjacency matrix $A$. When the
dynamics of node $i$ is affected by node $j$ via the coupling
function $h(x_i,x_j)$ with strength $g_{ij}$, $A_{ij}=1$ and a
link joins node $j$ to node $i$. Otherwise $A_{ij} = g_{ij} = 0$.
The coupling function satisfies $h_y(x,y) \equiv
\partial h(x,y)/\partial y \ne 0$ and $h_y >0$ thus
excitatory and inhibitory links have $g_{ij}>0$ and $g_{ij}<0$
respectively. We consider directed networks such that $A_{ij}$ and
$g_{ij}$ are generally asymmetric. We assume no self-interaction
so $A_{ii} \equiv 0$. External disturbance acting on node $i$ is
modeled by a Gaussian white noise $\eta_i$ of zero mean and
$\overline{ \eta_i(t)\eta_j(t')} = \sigma_i^2 \delta_{ij}
\delta(t-t')$. The overbar denotes ensemble average over different
realizations of the noise.

We consider weak noise so that we can linearize
Eq.~(\ref{network}) around its noise-free solution $X_i(t)$ to
obtain
\begin{equation}
\frac{d}{dt} {\bf \delta x} \approx Q {\bf \delta x} + {\bf \eta} \ , \label{deltax}
\end{equation}
where ${\bf \delta x}=(\delta x_1, \ldots, \delta x_N)^T$ and
$\delta x_i(t)= x_i(t) - X_i(t)$ is the deviation of the state
variable from the noise-free solution. The superscript $T$ denotes
a transpose and the  matrix $Q$, whose elements are given by
\begin{eqnarray}
\nonumber Q_{ij} &\equiv& g_{ij} A_{ij} h_y(X_i,X_j) \\
&& + \left[\sum_{k\ne i} g_{ik} A_{ik} h_x(X_i,X_k) +
f_i'(X_i)\right] \delta_{ij}  \ , \ \
\label{Qij}
\end{eqnarray}
contains information of the network structure. We
define $B(t_1,t_2)$ by
\begin{eqnarray}
\nonumber B(t_1,t_2) &\equiv& \overline{[{\bf
x}(t_1)-\overline{{\bf x}(t_1)}][{\bf x}(t_2)- \overline{{\bf
x}(t_2)}]^T}
\\
&=& \overline{[{\bf \delta x}(t_1)-\overline{{\bf \delta x}(t_1)}][{\bf \delta x}(t_2)-
\overline{{\bf \delta x}(t_2)}]^T}  \label{B}
\end{eqnarray}
where the last equality follows because
$\overline{X_i(t)}=X_i(t)$. $B_{ij}(t_1,t_2)$ is the covariance of
measurements of node $i$ at $t_1$ and  measurements of node $j$ at
time $t_2$. We focus on systems that approach a fixed point in the
noise-free limit so that $X_i$ and thus $Q$ are independent of
time. Using Eq.~(\ref{deltax}), we have~\cite{Arnold}
\begin{eqnarray}
\nonumber B(t_1,t_2) &\approx& e^{Qt_1}\{\overline{[{\bf
\delta x}(0)-\overline{{\bf \delta x}(0)}][{\bf \delta x}(0)- \overline{{\bf
\delta x}(0)}]^T}\}e^{Q^Tt_2} \\
&& \  + \int_0^{\min(t_1,t_2)} e^{Q(t_1-t')} D e^{Q^T(t_2-t')} dt'
 \ \label{resultB}
\end{eqnarray}
where $D$ is a diagonal matrix with $D_{ij}=\sigma_i^2 \delta_{ij}$. From
Eq.~(\ref{resultB}), we obtain
\begin{equation}
B(t+\tau,t) \approx e^{\tau Q} B(t,t) \qquad \tau > 0 \ ,
\label{newB}
\end{equation}
which is a noise-induced relation between the time-lagged
covariance matrix $B(t+\tau,t)$ and the covariance matrix
$B(t,t)$. For systems that have stationary dynamics, $B(t+\tau,t)$
depends on $\tau$ only, and can be approximated by (long) time
average:
\begin{equation}
B(t+\tau,t) \approx K( \tau) \equiv
\langle[{\bf x}(t+\tau)-\langle {\bf x}(t+\tau)\rangle][{\bf
x}(t)- \langle{\bf x}(t)\rangle]^T \rangle  \label{Ktau}
\end{equation}
where $\langle \cdots \rangle$ denotes a time average. Thus we
have
 \begin{equation}
e^{\tau Q} \approx B(t+\tau,t)B(t,t)^{-1} \approx K(\tau)K(0)^{-1}
\equiv e^{\tau M} \ . \label{new}
\end{equation}
It has been shown\cite{noise,PRE,PRErapid} that the presence of
noise leads to a one-to-one correspondence between network
structure and the inverse of covariance for undirected networks.
Equation~(\ref{new}) is a generalization of such result to
directed networks, which relates network structure to both the
time-lagged covariance of measurements taken at two different
times and covariance of measurements taken at the same time. The
matrix $M=\log[K(\tau)K(0)^{-1}]/\tau$ can be calculated from time
series measurements of $x_i(t)$. Equations~(\ref{Qij}) and
(\ref{new}) imply
\begin{equation}
M_{ij} \approx  g_{ij}A_{ij} h_y(X_i,X_j) \ , \qquad i \ne j
\label{result}
\end{equation}
hence the off-diagonal elements $M_{ij}$ would separate into
different groups depending on whether $A_{ij}=0$ or $A_{ij}=1$
(with $g_{ij}>0$ or $<0$). We identify these different groups of
$M_{ij}$ by clustering using Gaussian mixture model~\cite{note},
and, as a result,
reconstruct 
the links of the network.

In~\cite{SolvingPRE}, the authors focussed on the linearized
dynamics Eq.~(\ref{deltax}) and obtained a relation between the
``velocity-variable" covariance matrix between $dx_i/dt$ and
$x_j(t)$ and the covariance matrix. Their relation is the $\tau
\to 0$ limit of Eq.~(\ref{new}). To evaluate the velocity-variable
covariance matrix, one needs both $x_i(t)$ and $dx_i/dt$ but the
latter is usually not measured and is difficult to estimate
accurately from $x_i(t)$. Our method makes use of Eq.~(\ref{new}),
which holds for finite $\tau$, to reconstruct
 $A$ (and not $Q$) of directed networks using solely the measured $x_i(t)$.

We test our method using directed and weighted random (DWR) and
directed and weighted scale-free (DWSF)
networks~(Table~\ref{table0}). DWR1 and DWR2 are directed random
networks of connection probability 0.2 with DWR2 further
restricted to contain only unidirectional links such that
$A_{ij}=0$ for $i>j$. For both DWR1 and DWR2, $g_{ij}$'s are taken
from a Gaussian distribution $N(10,2)$ of mean 10 and standard
deviation 2 and all the links turn out to have $g_{ij}>0$. DWR1s
has the same adjacency matrix $A$ as DWR1 but $g_{ij}$'s are taken
from a different Gaussian distribution $N(10,10)$ such that the
links have both positive and negative $g_{ij}$. DWSF is
constructed by converting some bidirectional links of a undirected weighted scale-free
network~\cite{barrat2004}
into directional links with the power-law distribution of $g_{ij}$ kept
intact and the out-degree $k_{\rm out}(i)=\sum_j A_{ij}$
distribution being the same as the original power-law degree
distribution. We consider $f_i$ given by the logistic function
\begin{equation}
f_i(x)=r_i x(1-x) \label{logistic} \end{equation}
 and two coupling functions
 \begin{eqnarray}
h^{\rm diff}(x,y)&=&y-x \label{diffusive} \\
h^{\rm syn}(x,y)&=&(1/\beta_1)\{1+\tanh[\beta_2(y-y_0)]\} \ .
\label{syn}
\end{eqnarray} The linear diffusive coupling function
$h^{\rm diff}$ is a common model for gap junction coupling while
$h^{\rm syn}$ is a generalization~\cite{PRErapid} of a model for
synaptic coupling between neurons~\cite{Siam1990}. 
The parameters $\beta_1$, $\beta_2$, and $y_0$ are chosen such
that the steady-state values $X_i$ are close to $y_0$.

\begin{table}[tbh]
\centering
\begin{tabular}{|c|c|c|c|c|c|c|} \hline
Network & $N$ & $N_B$ & $N_U$ & $N_L$ & $\rho$ &  $g_{ij}$ \\
\hline
DWR1 & 100 & 186 & 1678 &  2050 & 0.207 & $N(10,2)$\\
DWR1s & 100 & 186 & 1678 &2050 & 0.207 & $N(10,10)$\\
DWR2 & 100 & 0 & 1035 & 1035 & 0.105& $N(10,2)$ \\
DWSF & 1000 & 3120 & 3730 & 9970 & 0.00998 & $P(g_{ij}) \sim
g_{ij}^{-6.6}$
\\ \hline
\end{tabular}
\caption{The networks studied. $N_B$ and $N_U$ are respectively
the number of bidirectionally- and unidirectionally-linked pairs
of nodes. The number of links $N_L$ is given by $2N_B+N_U$ and the
link density $\rho$ is equal to $N_L/[N(N-1)]$.
} \label{table0}
\end{table}

The theoretical basis of our method is given by Eq.~(\ref{new}),
which holds for general networked systems whose linearized
dynamics around the noise-free solution is described by
Eq.~(\ref{deltax}) with a time-independent $Q$. To investigate the
possible applicability of our method beyond such networks, we
consider three additional cases: (i) $f_i=0$ and $h^{\rm
cubic}(x,y) = (y-x)^3$, (ii) networks whose nodes have
two-dimensional state variables $(x_i(t),y_i(t))$ described by the
nonlinear FitzHugh-Nagumo (FHN) dynamics~\cite{FHN}, which is a
common model for neurons,
\begin{eqnarray} \label{FHN1}
\dot{x_i}&=& 
(x_i-{x_i^3}/{3}-y_i)/\epsilon +\displaystyle\sum\limits_{j\ne i}
g_{ij} A_{ij}h(x_i,x_j)+\eta_i \qquad
\\
\dot{y_i}&=&x_i+\alpha \,  \label{FHN2}
\end{eqnarray}
with $\epsilon=0.01$  and (iii)
networks whose nodes have three-dimensional state variables
$(x_i(t),y_i(t),z_i(t))$ described by the nonlinear R\"ossler
dynamics~\cite{Rossler}:
\begin{eqnarray}
\dot{x}_i
&=& -y_i -z_i  +\sum_{j\ne i} g_{ij}A_{ij} h(x_i,x_j) +\eta_i \\
\dot{y}_i &=& x_i + a y_i +\sum_{j\ne i} g_{ij}A_{ij}h(y_i,y_j)
\\
\dot{z}_i &=& b + z_i(x_i-c) +\sum_{j\ne i} g_{ij}A_{ij}h
(z_i,z_j) \  \qquad
\end{eqnarray}
with $a=b=0.2$ and $c=9$ (for such parameters, the dynamics is chaotic if the nodes are decoupled). In case (i), the noise-free solution is
$X_i(t)=X_0$ thus $h^{\rm cubic}_y(X_i,X_j)=0$ and hence the
linearized dynamics cannot be a good
approximation. In case (iii) the noise-free solution depends on
time.  We integrate the equations of motion using the
Euler-Maruyama method, then calculate $K(\tau)$ and $K(0)^{-1}$
from $x_i(t)$ with an average over $T_{av}$ to obtain $M$. We use
only $x_i(t)$ for reconstruction even for cases (ii) and (iii) and
take $\tau=5 \times 10^{-4}$, which is the same as the sampling
interval of $x_i(t)$, $\sigma_i=1$, and $T_{av}=1000$ unless
otherwise stated.

\begin{figure}[tbh]
\centering
\subfigure[]{\includegraphics*[angle=270,width=3.80cm]{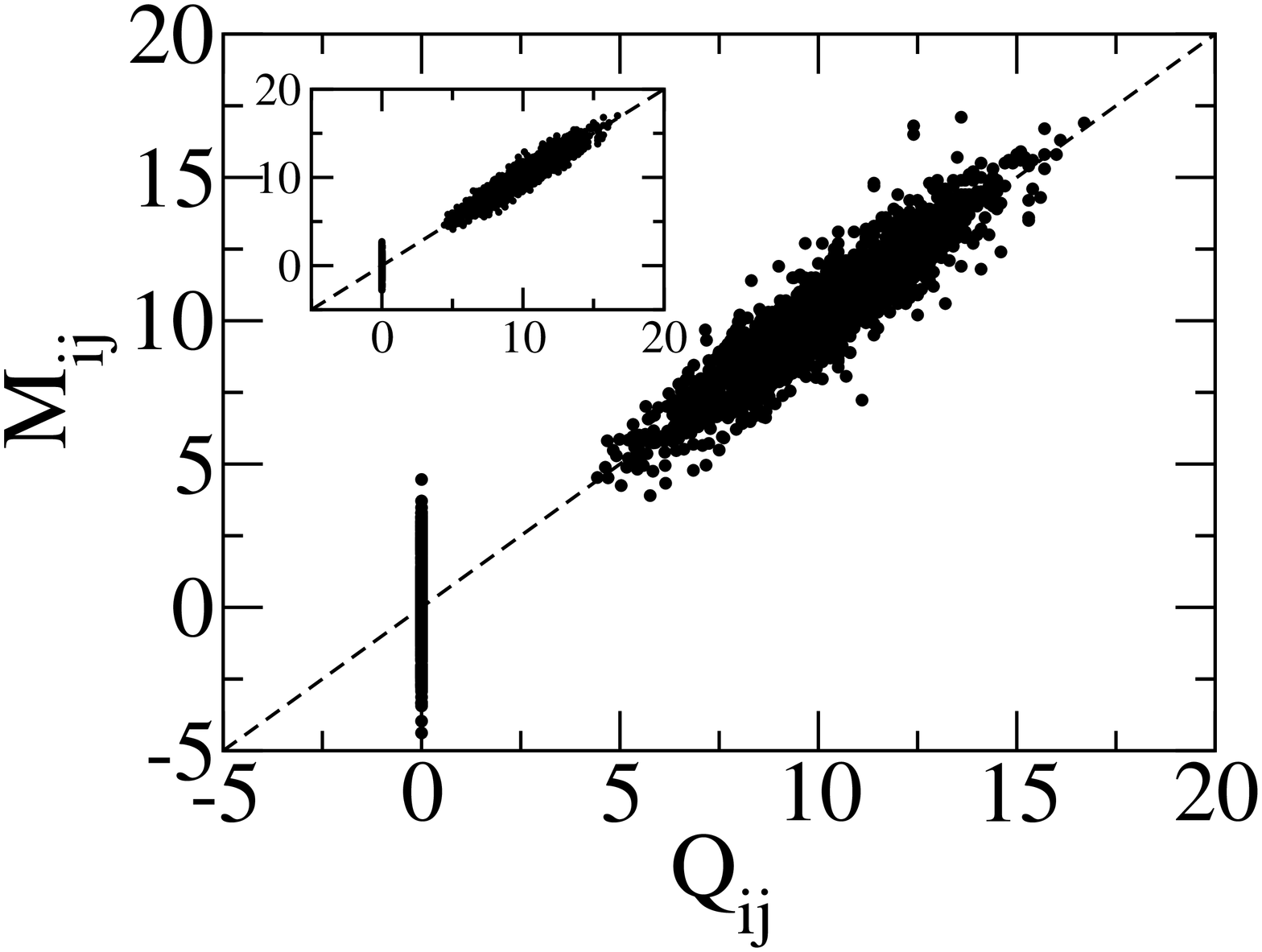}}
\subfigure[]{\includegraphics*[angle=270,width=3.80cm]{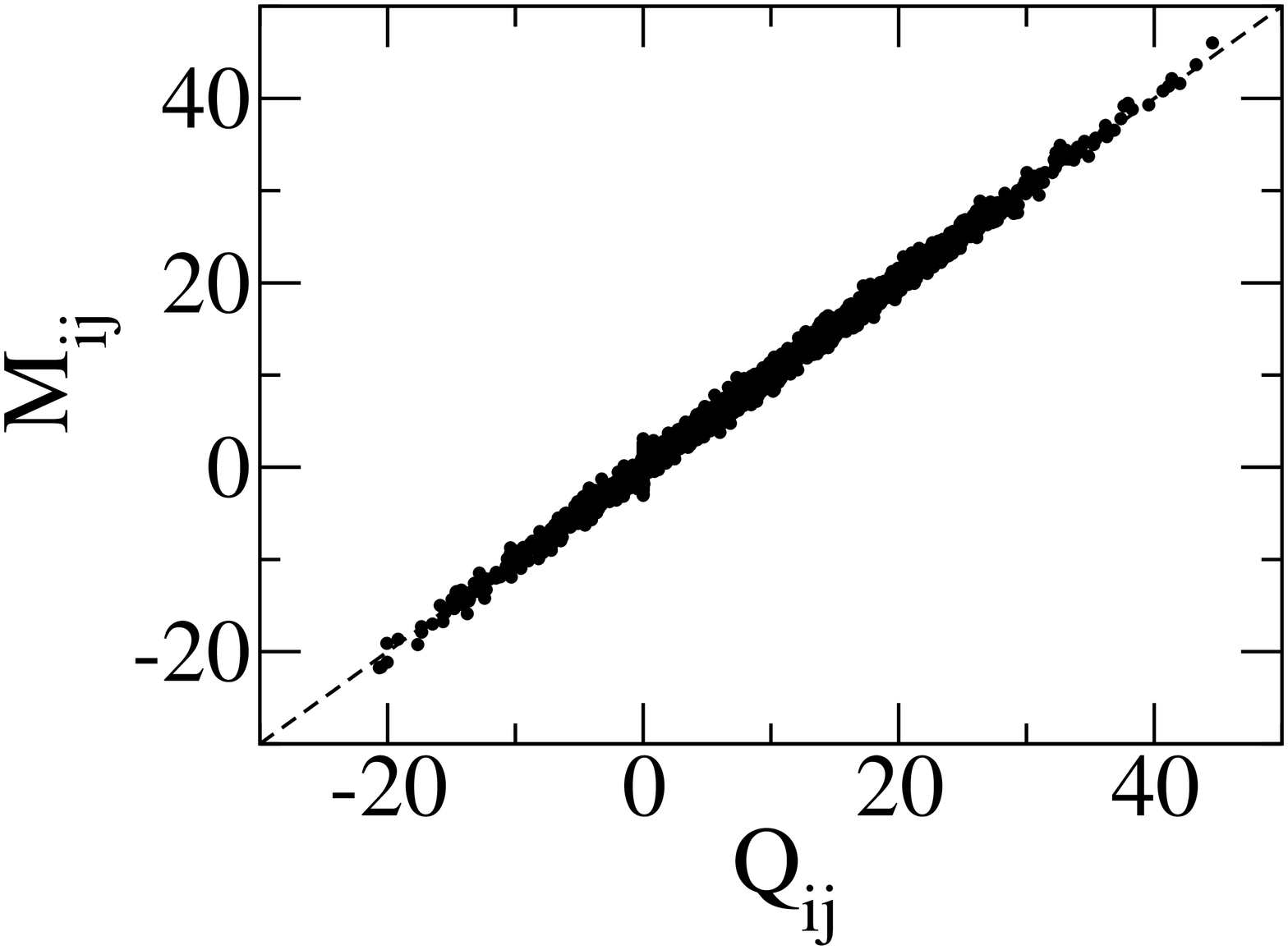}}
\subfigure[]{\includegraphics*[angle=270,width=3.80cm]{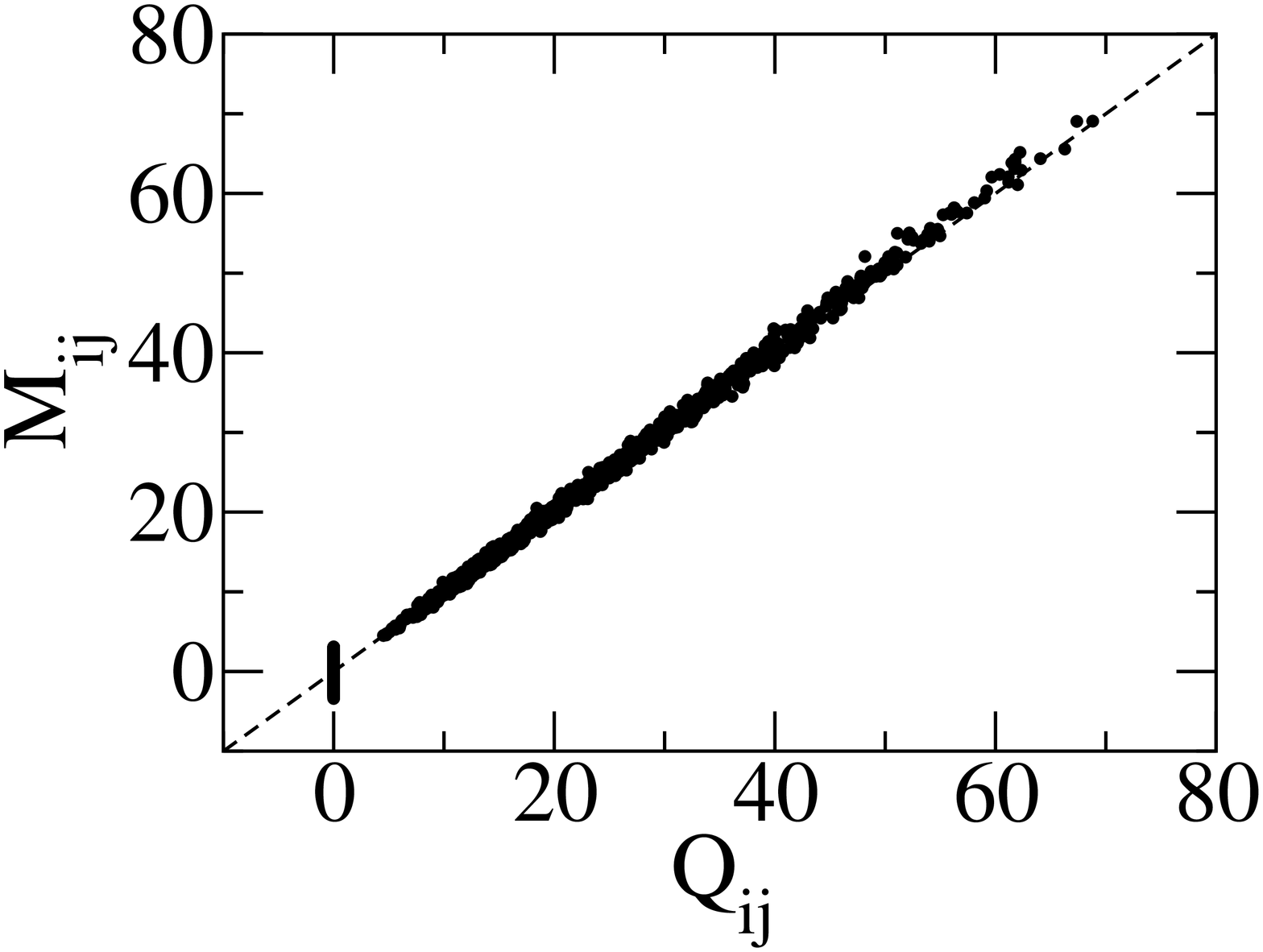}}
\subfigure[]{\includegraphics*[angle=270,width=3.80cm]{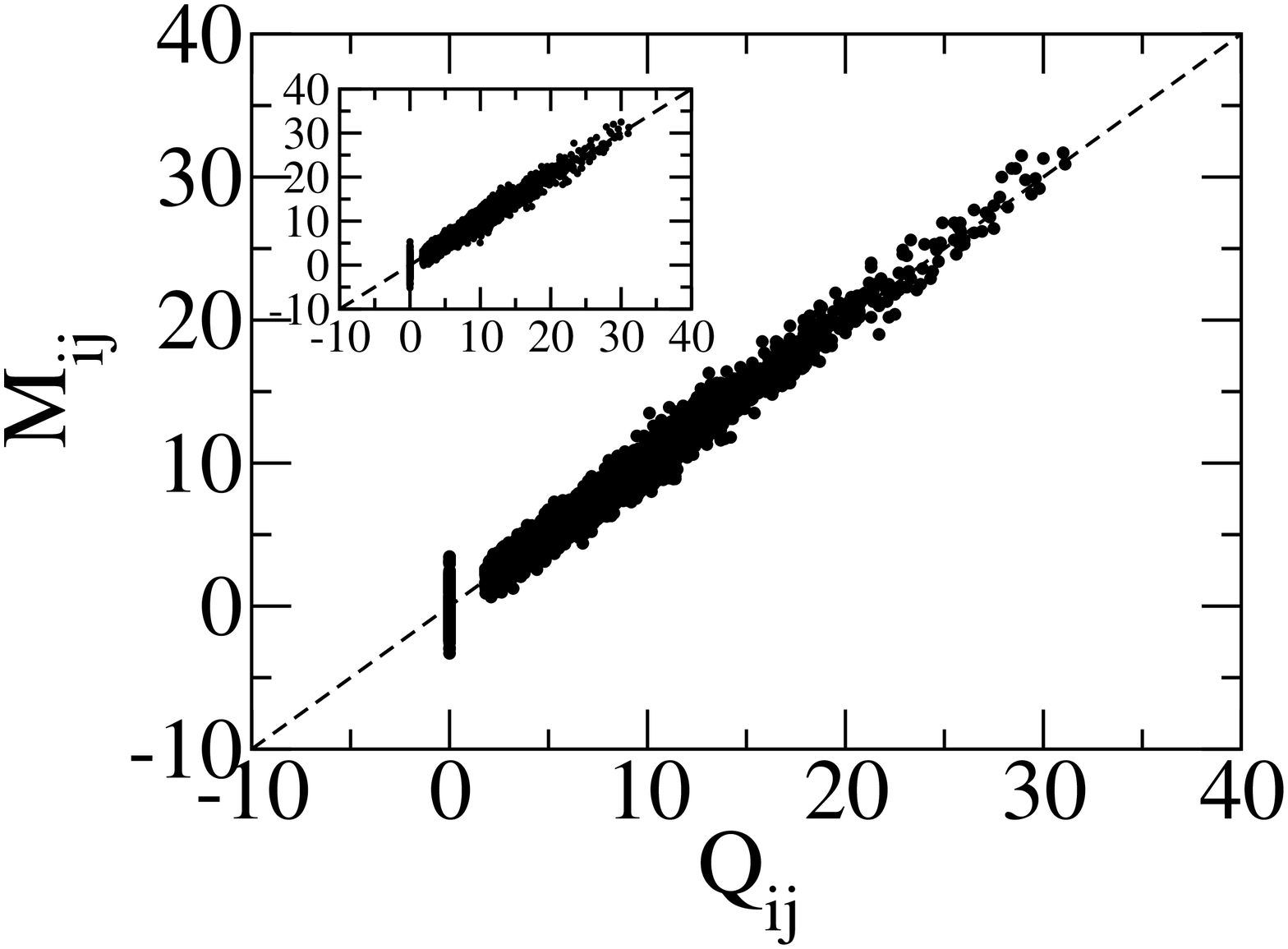}}
\subfigure[]{\includegraphics*[angle=270,width=3.80cm]{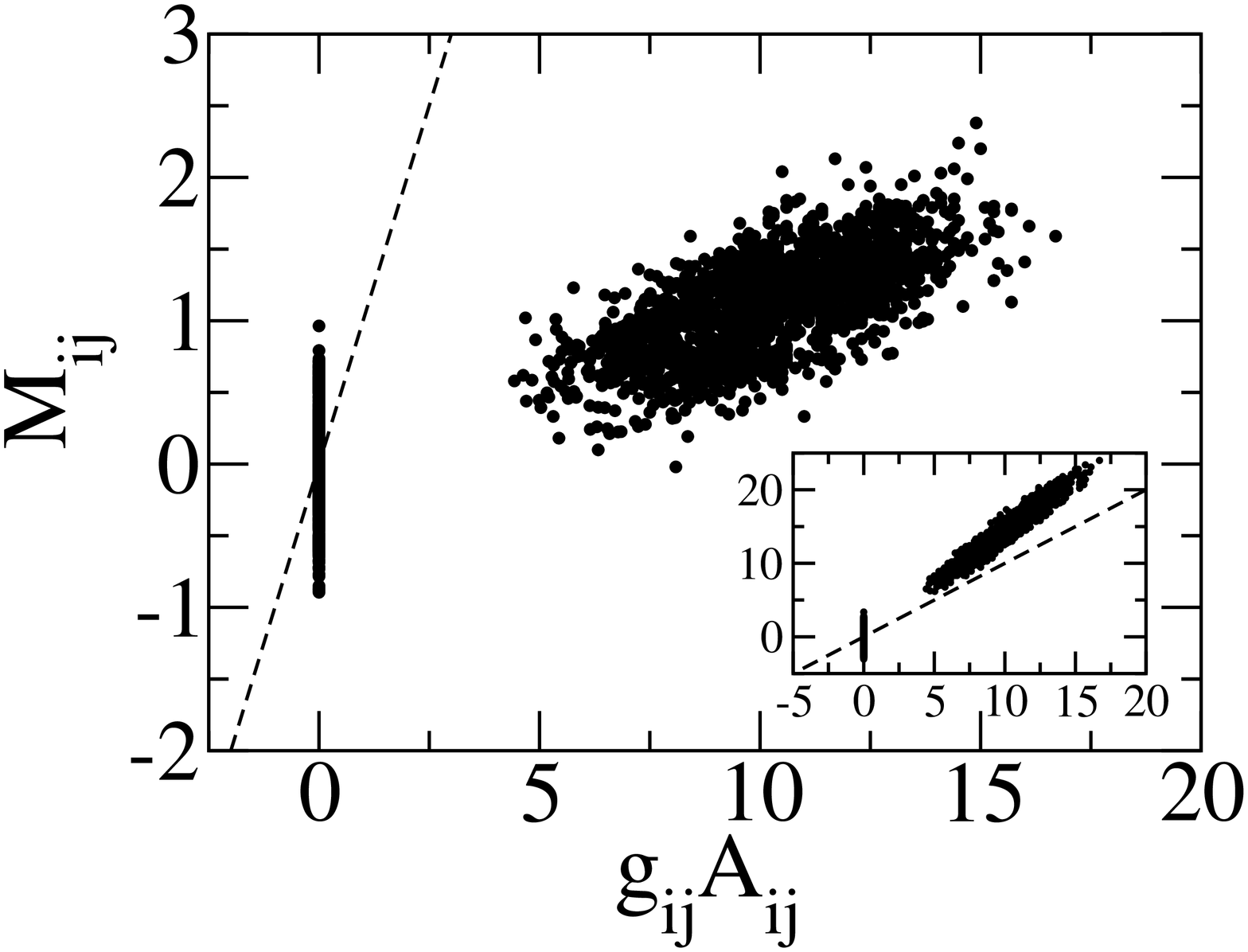}}
\subfigure[]{\includegraphics*[angle=270,width=3.80cm]{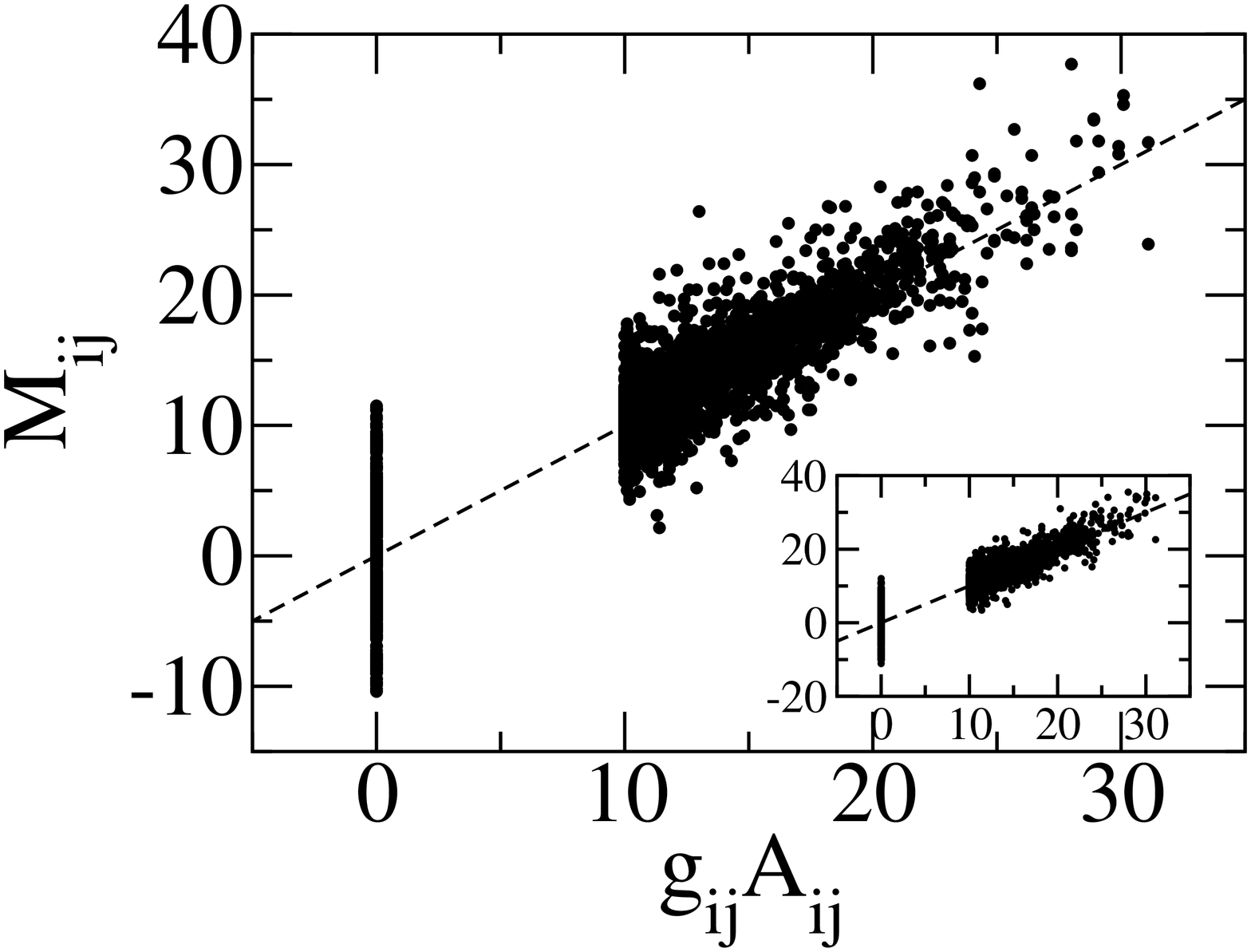}}
 \caption{$M_{ij}$ versus $Q_{ij}=g_{ij}A_{ij}h_y(X_i,X_j)$ or
 $g_{ij}A_{ij}$
 for (a) DWR1 with nonuniform noise (case 2) and inhomogeneous $f_i$ (case 3) (inset),
(b) DWR1s with both excitatory and inhibitory links (case 5), (c)
DWR2 with logistic $f_i$ and $h^{\rm syn}$ (case 7), (d) DWSF with
logistic $f_i$ and $h^{\rm syn}$ with $T_{av}=2000$ (case 9) and
$T_{av}=1000$ (inset), (e) DWR1 with $h^{\rm cubic}$ (case 10) and
FHN with $h^{\rm syn}$ (case 12, inset), and (f) DWSF with R\"{o}ssler (case 17)
and FHN (case 16, inset) dynamics. See
Table~\ref{table1} for detailed descriptions of the different
cases. Dashed line is $y=x$.}
 \label{fig1}
 \end{figure}

We compare the off-diagonal elements of $M$ and $Q$ in
Fig.~\ref{fig1}. The data points scatter around the line $y=x$,
confirming Eq.~(\ref{result}). One cause for the data scatter is
due to the finite sample size, the noise generated in our
simulations is not exactly delta-correlated in time. [For noise
with finite correlation time, Eq.~(\ref{new}) would be modified to
$e^{\tau M} \approx e^{\tau Q} + E$ with $M \approx Q$ with an
error related to $E$.] This data scatter is reduced when the
sample size or $T_{av}$ is increased~(see Fig.~\ref{fig1}d). For
the additional cases (i)-(iii), we compare $M_{ij}$ and
$g_{ij}A_{ij}$ and find
\begin{equation} M_{ij} \approx C g_{ij}A_{ij} \ , \qquad i \ne j
\label{ext}
\end{equation} for some constant $C$. This interesting result
indicates that our method can also be applicable to these cases.

\begin{figure}[tbh]
\centering
\subfigure[]{\includegraphics*[angle=270,width=4.2cm]{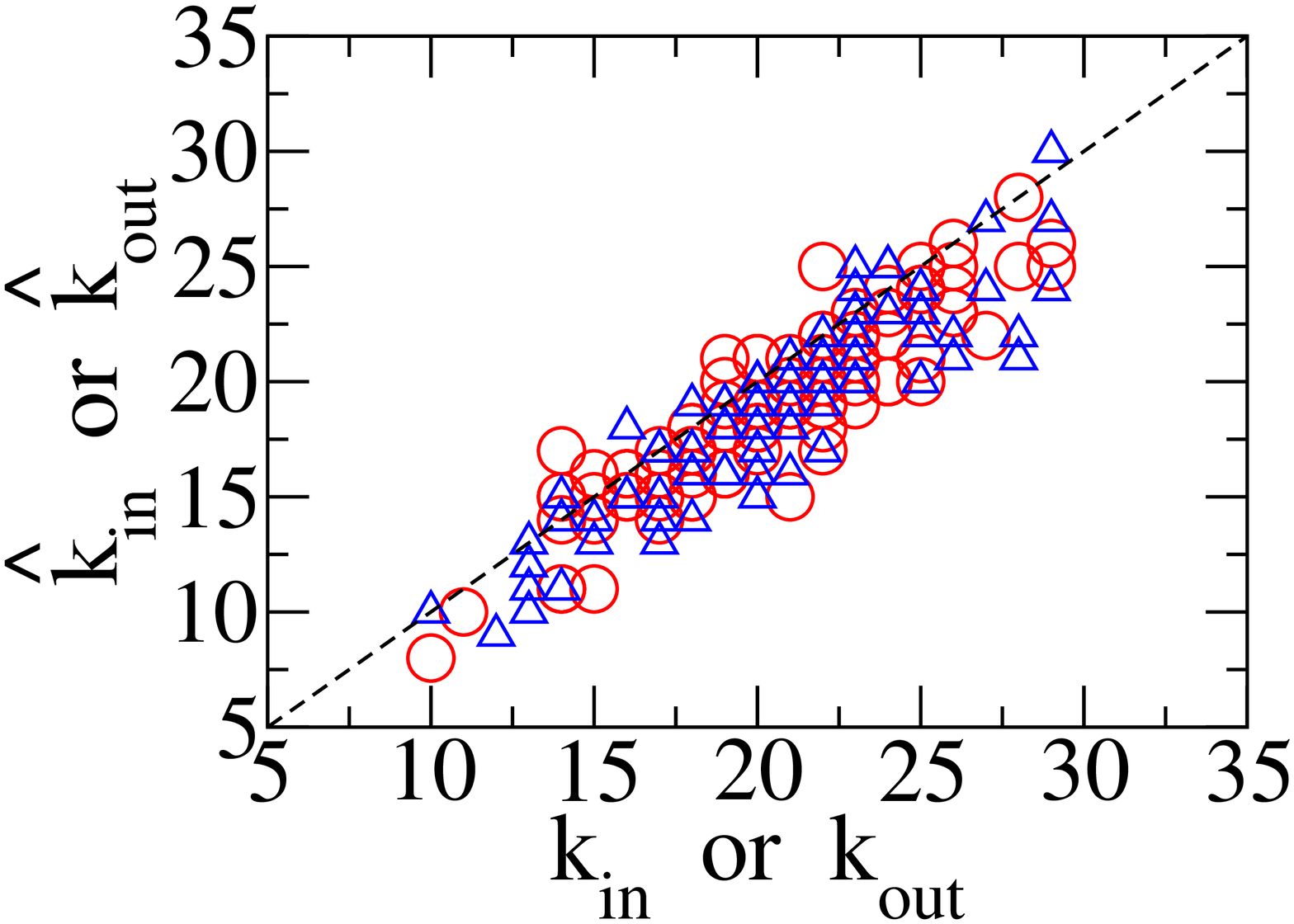}}
\subfigure[]{\includegraphics*[angle=270,width=4.2cm]{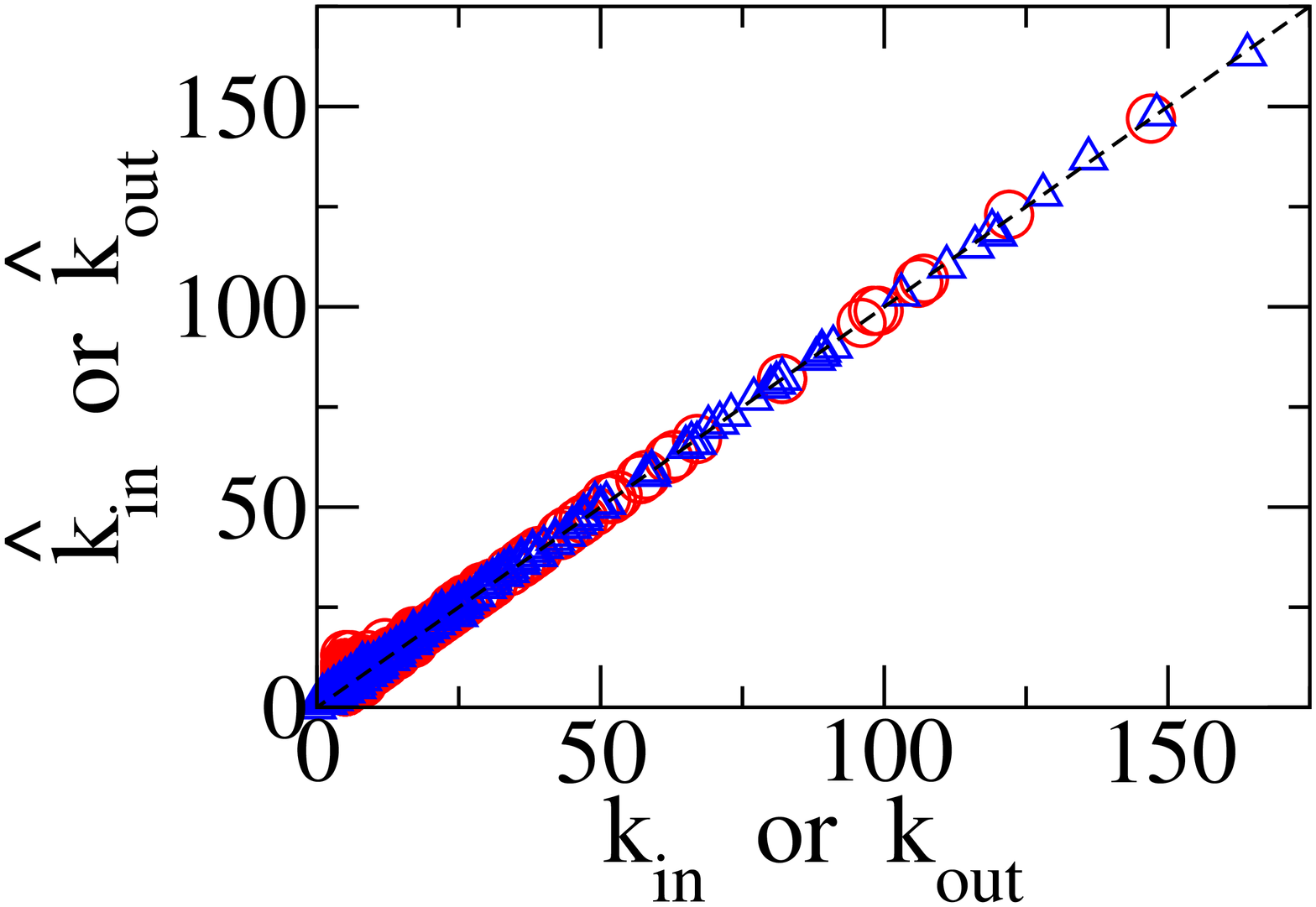}}
\caption{Comparison of the reconstructed in-degrees $\hat{k}_{\rm
in}$~(triangles) and out-degrees $\hat{k}_{\rm out}$~(circles)
with the actual values
 for (a) case 5 and
 (b) case 9. }
 \label{fig2}
 \end{figure}

It is common to measure the accuracy of a method by its
sensitivity and specificity. However, for sparse networks with
link density $\rho \ll 1$, the number of incorrectly inferred links
can be substantial leading to a greatly
distorted reconstructed network
even when specificity is close to 1. So, we
measure the accuracy instead by the error rates ${\rm FN}/N_L$ and
${\rm FP}/N_L$, which are respectively the proportion
of the links that are missed (false negatives FN) and the ratio
of incorrectly inferred links (false positives FP) to the
number of links~(see Table~\ref{table1}). Sensitivity is given by
$1-{\rm FN}/N_L$ while specificity is given by $1-({\rm
FP}/N_L)\rho/(1-\rho)$.  The two error rates are less than 10\%
for all the cases studied (for case 9, the error rates are lowered
to less than 10\% with a longer $T_{av} =2000$), including network
acted upon by nonuniform noise~(case 2), network with
inhomogeneous intrinsic dynamics~(case 3), network with both
excitatory and inhibitory links~(case 5), and networks with
dynamics given by the additional cases (i)-(iii) beyond the
description given by Eq.~(\ref{deltax})~(cases 10-17). From
Eq.~(\ref{result}) or the approximate extension Eq.~(\ref{ext}),
we expect that weak links with small $|g_{ij}|$ are difficult to
detect. Indeed all links that are missed are of relatively weak
coupling strength. For the cases that have the highest
error rates in DWR and DWSF networks respectively, we compare directly the
reconstructed in- and out-degree of the nodes, $\hat{k}_{\rm in}$
and $\hat{k}_{\rm out}$ with the actual values. Good agreement is
found as shown in Fig.~\ref{fig2}. Our method can thus capture the
power-law out-degree distribution of DWSF rather well.


\begin{table}[tbh]
\centering
\begin{tabular}{|c|c|c|c|c|c|}
\hline Case & Network & Dynamics & ${\rm FN}/{N_L}$ & ${\rm FP}/{N_L}$ & $e_G$   \\
\hline
1 & DWR1 & logistic $r_i=10$;  $h^{\rm diff}$& 0 & 0 & 5.7 \\
2 & DWR1 & logistic $r_i=10$; $h^{\rm diff}$ & 0 & 0.63 & 6.0\\
3 & DWR1 & logistic; $h^{\rm diff}$ & 0 & 0 & 6.0 \\
4& DWR1 & logistic $r_i=10$; $h^{\rm syn}$ & 2.98& 2.00  & 14.2\\
5 & DWR1s & logistic $r_i=10$; $h^{\rm diff}$ & 9.32 & 1.90 &9.9 \\
6 & DWR2 & logistic $r_i=50$; $h^{\rm diff}$ & 0 & 0 & 4.3\\
7 & DWR2 & logistic $r_i=50$; $h^{\rm syn}$ & 0& 3.38 &4.5 \\
8 & DWSF & logistic $r_i=100$; $h^{\rm diff}$ & 0.50 & 1.11 &6.9\\
9& DWSF & logistic $r_i=100$; $h^{\rm syn}$ & 0.65& 2.84 & 10.3 \\
\hline
10& DWR1 & $f_i=0$; $h^{\rm cubic}$ & 4.24& 2.78 & 16.6 \\
11& DWR1 & FHN $\alpha=1.05$;  $h^{\rm diff}$ & 0 & 0 &5.7\\
12& DWR1 & FHN $\alpha=2$; $h^{\rm syn}$ & 0 & 0 & 4.9\\
13& DWR1 & R\"{o}ssler; $h^{\rm diff}$ & 0 & 0 & 6.9\\
14& DWR2 & FHN $\alpha=1.05$; $h^{\rm diff}$ & 0 & 0 & 3.3\\
15& DWR2 & R\"{o}ssler; $h^{\rm diff}$ & 0 & 0.10 & 5.4\\
16& DWSF & FHN $\alpha=1.05$; $h^{\rm diff}$ & 0.32 & 1.20 & 5.8\\
17& DWSF & R\"{o}ssler; $h^{\rm diff}$ & 0.35 & 1.58 & 6.2\\
\hline\end{tabular} \caption{Accuracy of our reconstruction as
measured by the error rates FN$/N_L$ and FP$/N_L$ (in \%) for the
cases studied. Case 2:~$\sigma_i$ taken from $N(1,0.2)$. Case
3:~$r_i$ taken from a uniform distribution $U(1,50)$ from 1 to 50.
Case 4:~$(\beta_1,\beta_2,y_0)=(2,0.5,4)$. Case
7:~$(\beta_1,\beta_2,y_0)=(0.1,0.5,4)$. Case
9:~$(\beta_1,\beta_2,y_0)=(0.5, 0.5, 4)$ and $T_{av}=2000$. Case
12:~ $(\beta_1,\beta_2,y_0)=(0.1,2,-1)$ and $\sigma_i=0.1$. Cases
14 and 16:~$\sigma_i=0.1$. $e_G$ is the average percentage error
of the reconstructed relative coupling strength $\hat{G}_{ij}$ or
$\hat{G}^{\rm out}_j(i)$. } \label{table1}
\end{table}

\vspace{0.4cm}

\vspace{0.5cm}

\begin{figure}[tbh]
\vspace{0.2cm} \centering
\subfigure[]{\includegraphics*[angle=270,width=3.80cm]{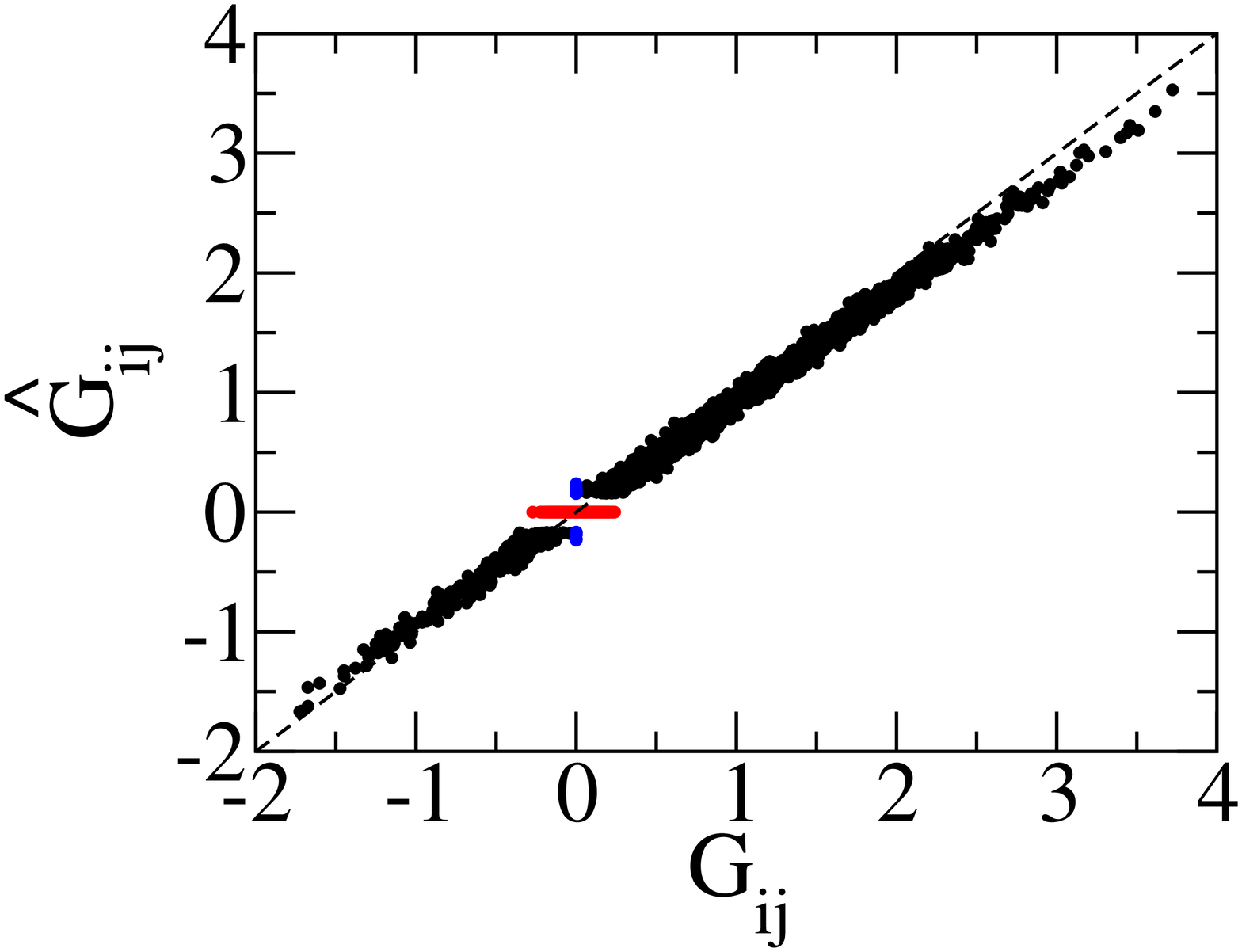}}
\subfigure[]{\includegraphics*[angle=270,width=3.80cm]{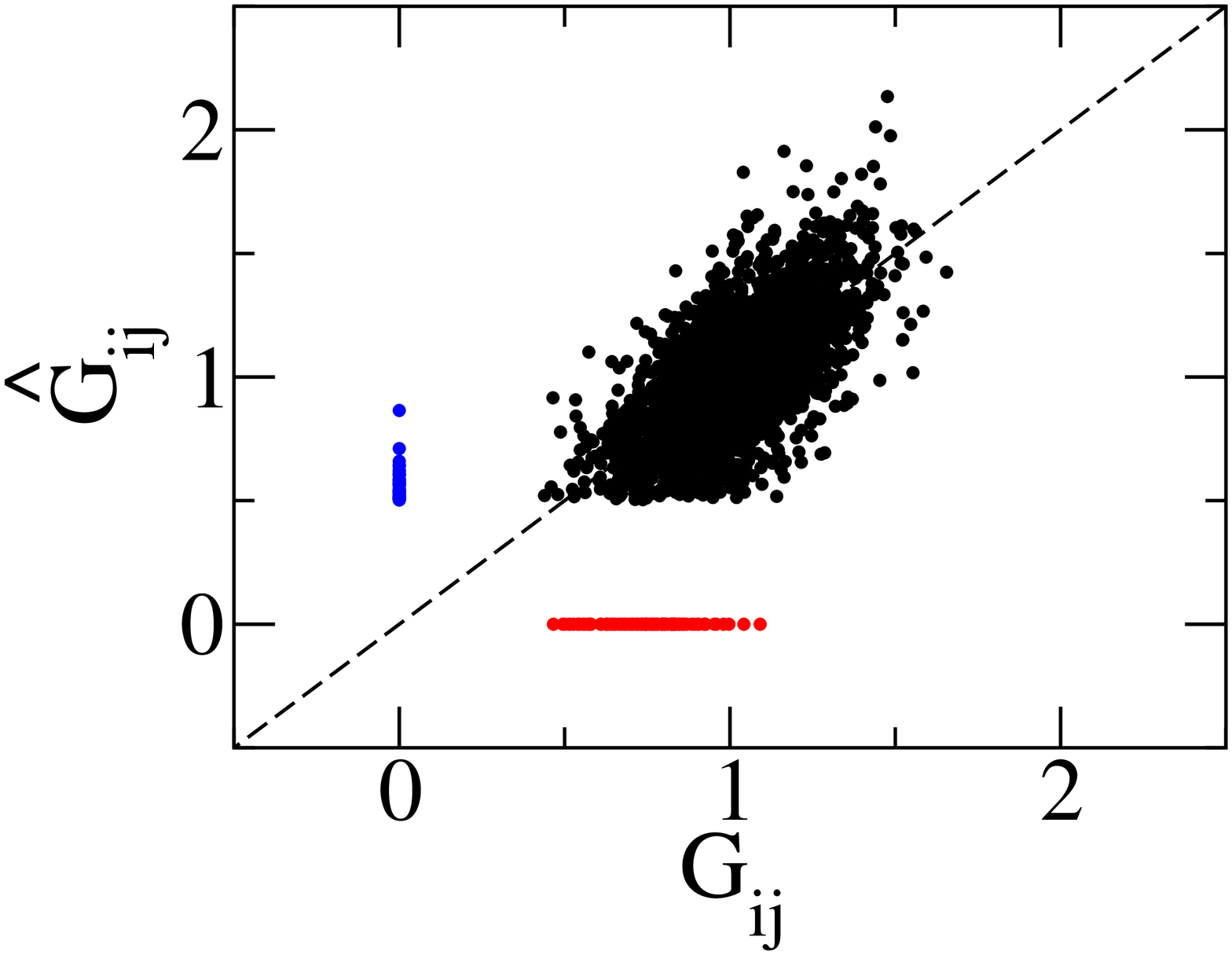}}
\subfigure[]{\includegraphics*[angle=270,width=3.80cm]{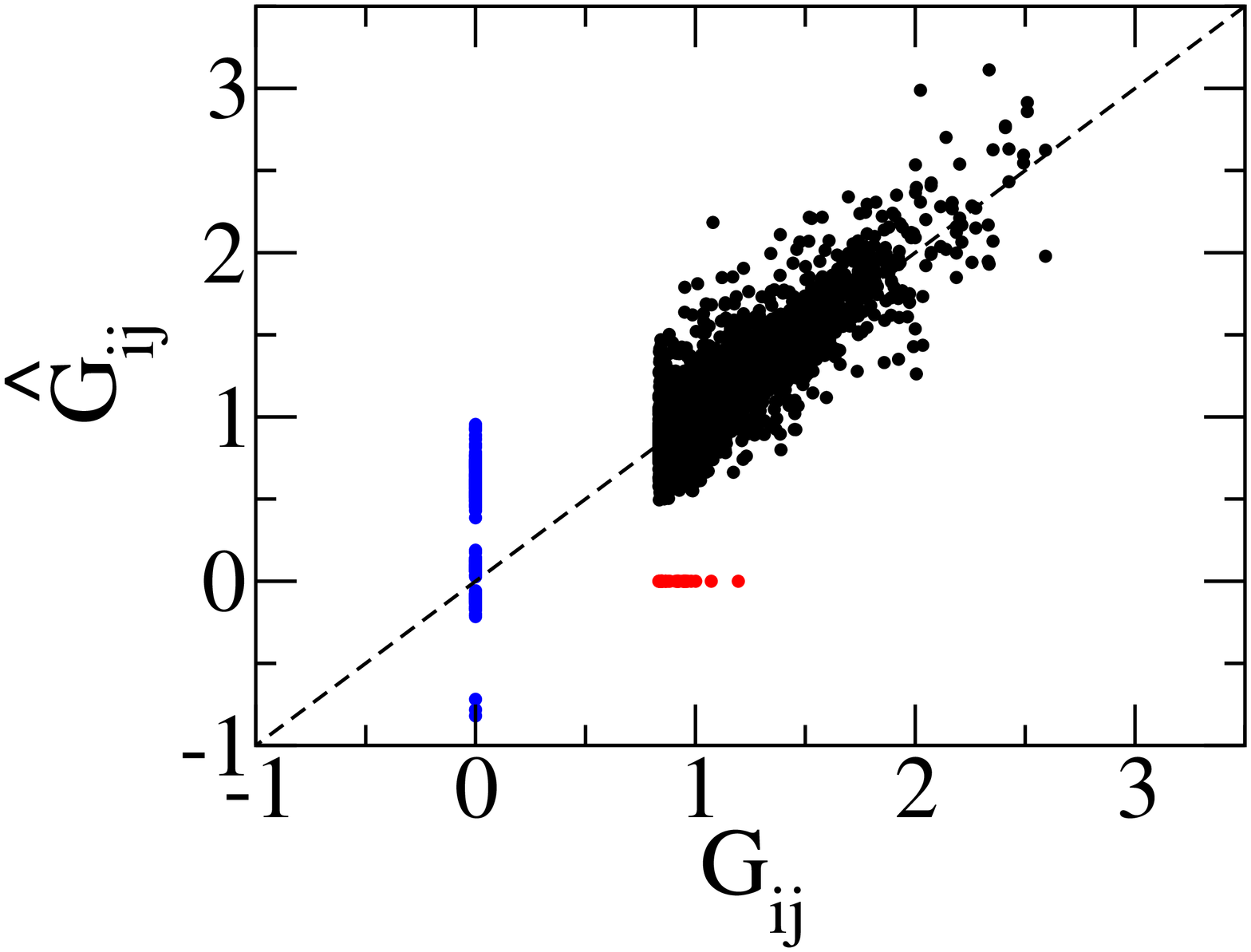}}
\subfigure[]{\includegraphics*[angle=270,width=3.80cm]{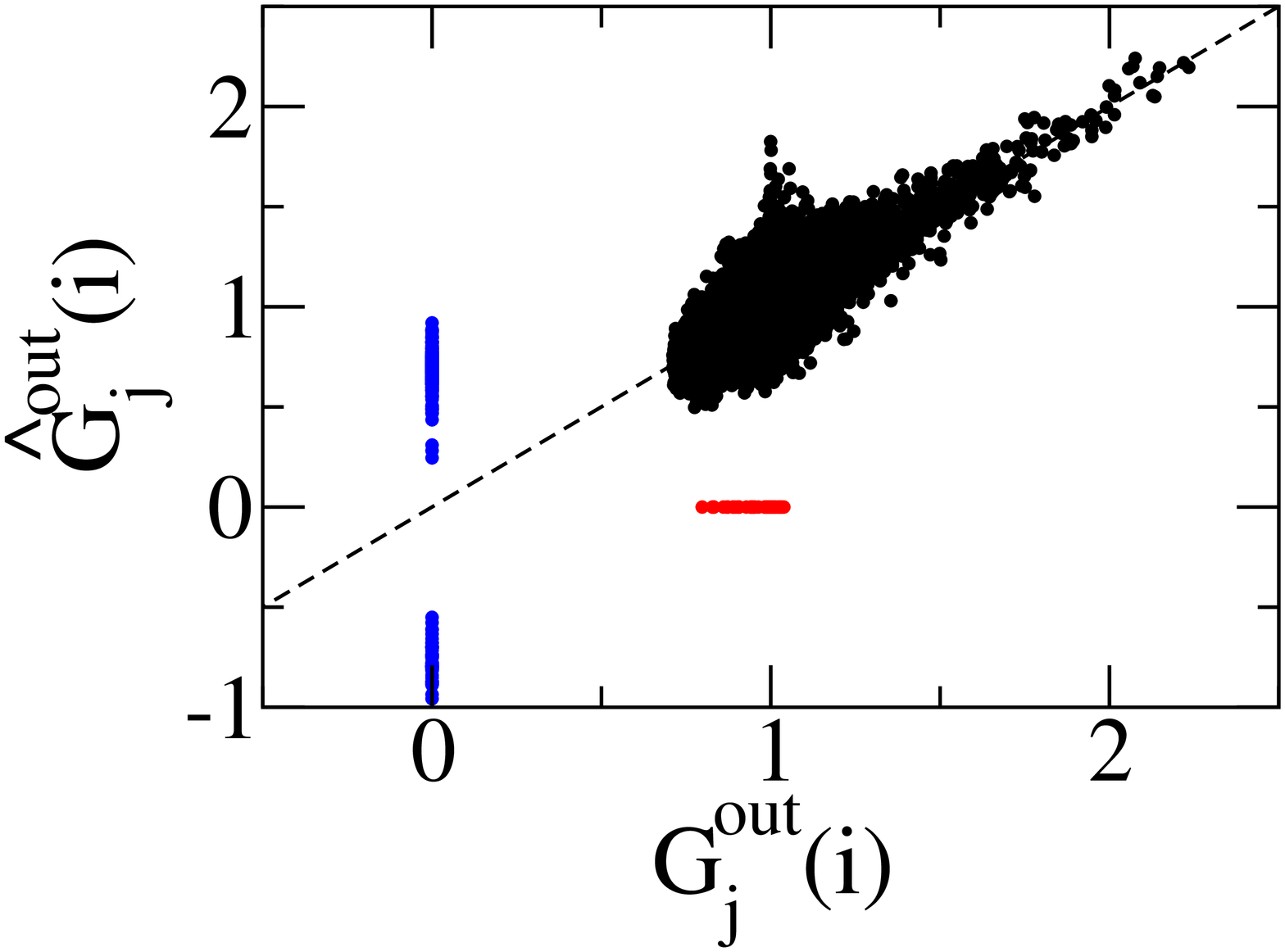}}
\caption{Color online: Comparison of reconstructed
 $\hat{G}_{ij}$ and $\hat{G}^{\rm
out}_j(i)$ with the actual values for (a) case 5, (b) case 10, (c)
case 17, and (d) case 9. Missed links (red) have $G_{ij}\ne 0$ or
${G}^{\rm out}_j(i)\ne0$ but $\hat{G}_{ij}=0$ or $\hat{G}^{\rm
out}_j(i)=0$ while incorrectly inferred links (blue) have
$G_{ij}=0$ or ${G}^{\rm out}_j(i)=0$ but $\hat{G}_{ij}\ne 0$ or
$\hat{G}^{\rm out}_j(i)\ne 0$. Dashed line is $y=x$.} \label{fig3}
\end{figure}

 We define
the relative coupling strength of an in- and out-link of node $i$
from and to node $j$ by \begin{equation}
G^{\rm in}_{j}(i)  \equiv \frac{g_{ij}}{\langle g\rangle_{\rm
in}(i)} \ ; \qquad  G^{\rm out}_{j}(i)  \equiv
\frac{g_{ji}}{\langle g\rangle_{\rm out}(i)} \
\end{equation}
respectively, where $\langle g \rangle_{\rm in}(i) \equiv {\sum_j
|g_{ij}|A_{ij}}/{k_{\rm in}(i)}$ and $\langle g \rangle_{\rm
out}(i) \equiv {\sum_j |g_{ji}|A_{ji}}/{k_{\rm out}(i)}$
 are the average (absolute) coupling strength of  the in- or  out-links of node
$i$. If $h_y(x,y)$ depends on $y$ only as for $h^{\rm syn}$,
Eq.~(\ref{result}) implies
\begin{equation} G^{\rm out}_{j}(i)   \approx \frac{M_{ji}\hat{k}_{\rm
out}(i)}{\sum_{k \leftarrow i} |M_{ki}| } \equiv \hat{G}^{\rm
out}_{j}(i) \  {\rm if} \ h_y = h_y(y) \ . \  \label{Gout}
\end{equation}
$\sum_{k \leftarrow i}$ represents a sum over nodes $k$ that
are reconstructed  to be linked from node $i$. Similarly if
$h_y(x,y)$ depends on $x$ only, we can reconstruct $G^{\rm
in}_j(i)$ using $\hat{G}^{\rm in}_{j}(i) \equiv
{M_{ij}\hat{k}_{\rm in}(i)}/{\sum_{k \rightarrow i} |M_{ik}| }$,
where $\sum_{k \rightarrow i}$ represents a sum over nodes $k$
that are reconstructed to link to node $i$.
 If $h_y(x,y)$ is a constant as for $h^{\rm diff}$, Eq.~(\ref{result}) gives
\begin{equation}
G_{ij} \equiv \frac{g_{ij}}{\langle g \rangle} \approx
\frac{M_{ij} \hat{k}_{\rm tot}}{{\sum_{n,l \leftrightarrow n}
|M_{nl}|}} \equiv \hat{G}_{ij} \label{Gij}
\end{equation}
where $\langle g \rangle \equiv \sum_{ij} |g_{ij}|
A_{ij}/\sum_{ij} A_{ij}$ and $\hat{k}_{\rm tot} =\sum_i
\hat{k}_{\rm in}(i)=\sum_i \hat{k}_{\rm out}(i)$. The extension
Eq.~(\ref{ext}) implies that Eq.~(\ref{Gij}) should hold
approximately for the additional cases (i)-(iii). In
Fig.~\ref{fig3}, we compare the reconstructed relative coupling
strength $\hat{G}_{ij}$ and $\hat{G}^{\rm out}_j(i)$ with the
actual values. The average percentage error (excluding missed and
incorrectly predicted links) ranges between 3.3\% to 17\% for all
the cases studied~(see Table~\ref{table1}).

In conclusion, we have presented a method that reconstructs links
in directed networks. Our method makes use of a noise-induced
relation Eq.~(\ref{new}) that gives a one-to-one correspondence of
the network structure and both the time-lagged covariance and
covariance of measurements. Using numerically simulated data, we
have shown that our method can successfully reconstruct the
network structure with low error rates for DWR and DWSF networks
with different nonlinear dynamics and coupling functions. Our
method is general and requires only time series measurements of
the nodes. For coupling functions that have additional properties,
our method can further reconstruct the weights of the
links.

\begin{acknowledgments} We acknowledge the Hong Kong Research Grants Council (grant no. CUHK 14300914) for
support, and thank K.C. Lin for introducing us to
clustering analysis using Gaussian mixture models in MATLAB.
\end{acknowledgments}


\begin{thebibliography}{00}
\bibitem{note} We use MATLAB fitgmdist to fit the $M_{ij}$ data
by a Gaussian mixture model of 2 components then use MATLAB
cluster to obtain the posterior probabilities of each component
for each data point. For each data point $M_{ij}$, if the
posterior probability of the component with a mean closest to 0
(which is associated with the component for unconnected nodes) is
less than 0.5, then this $M_{ij}$ is inferred to be from
connected nodes and $\hat{A}_{ij}=1$. Otherwise, $\hat{A}_{ij}=0$.
For DWR networks, clustering analysis is performed on all $M_{ij}$
($i\ne j$) values as a whole while for DWSF network, clustering
analysis is performed on $M_{ij}$ values for fixed $j$ separately
for each $j$.
\end{thebibliography}
\end{document}